\documentstyle[pasjskel,graphicx,natbib,twocolumn]{article}

\begin{document}

\title{Kilo-Second Quasi-Periodic Oscillations in the Cataclysmic
Variable DW Cancri}

\author{Makoto \textsc{Uemura}, Taichi \textsc{Kato}, and 
Ryoko \textsc{Ishioka},}
\affil{Department of Astronomy, Faculty of Science, Kyoto University,
  Sakyou-ku Kyoto 606-8502}
\email{uemura@kusastro.kyoto-u.ac.jp}
\author{Rudolf \textsc{Novak},}
\affil{Nicholas Copernicus Observatory, Krav\'{\i} hora 2, Brno
  616 00, Czech Republic}
and
\author{Jochen \textsc{Pietz}}
\affil{Rostocker Str. 62, 50374 Erftstadt, Germany}


%


\begin{abstract}

 Our photometric monitoring revealed that DW Cnc, which was originally
 classified as a dwarf nova ($V=15$--$17.5$), remained at a bright state
 of $R_{\rm c}=14.68\pm 0.07$ for 61 days.  In conjunction with optical
 spectra lacking a strong He~{\sc ii} emission line, we propose that the
 object is not a dwarf nova, but a non-magnetic nova-like variable.
 Throughout our monitoring, the object showed strong quasi-periodic 
 oscillations (QPOs) with amplitudes reaching about 0.3 mag.  Our period 
 analysis yielded a power spectrum with two peaks of QPOs, whose center 
 periods are $37.5\pm 0.1$ and $73.4\pm 0.4$ min and, furthermore, with 
 a significant power in frequencies lower than the QPOs.  DW Cnc is a 
 unique cataclysmic variable in which kilo-second QPOs were continuously 
 detected for 61 days.  We propose two possible interpretations of DW
 Cnc: (i) A permanent superhumper below the period minimum of
 hydrogen-rich cataclysmic variables. (ii) A nova-like variable having
 an orbital period over 3 hours.  In this case, the QPOs may be caused
 by trapped disk oscillations.

\end{abstract}

\section{Introduction}

Cataclysmic variables (CVs) are semi-detached binaries which contain 
a white dwarf and a late-type secondary star (\cite{war95}).  
The overflowing gas from the Roche lobe-filling secondary
star forms an accretion disk, which plays a key role in various
activities of CVs.  From the photometric and spectroscopic
characteristics, the systems are classified with some sub-groups, for
example, classical novae, dwarf novae (DNe), nova-like variables (NLs), 
and magnetic CVs of polars and intermediate polars (IPs).  

Quasi-periodic oscillations (QPOs) have been detected in CVs on a
variety of time scales.  Rapid QPOs with a period of a few seconds have 
been observed in some magnetic CVs.  They are considered to 
originate from the shock region between a white dwarf and accretion
streams (Langer et al. 1981, 1982).  During DN outbursts, we know
three types of temporary short-term modulations, that is, dwarf nova
oscillations (DNOs), QPOs, and super-QPOs.  
DNOs are highly coherent modulations with typical periods of a few tens of 
seconds.  The period of QPOs during DN outbursts is generally longer than
that of DNOs ($P_{\rm QPO}=30$--$1000$ s).  DNOs and QPOs have been
reported also in some NLs and classical novae (\cite{pat81a}).  The 
origin of DNOs has been proposed to be a white dwarf (\cite{war72}; 
\cite{pap78}), an inner edge of the disk (\cite{bat73}) or a boundary 
layer (\cite{oku92}; \cite{col00}); however, it is still an issue.  On
the other hand, the oscillation of a disk or a boundary layer has been 
suggested for QPOs (\cite{oku92}; \cite{col00}).  While the amplitudes
are quite small ($\sim 0.05\; {\rm mag}$) in DNOs and QPOs, super-QPOs 
have higher amplitudes, reaching $\sim 0.2\; {\rm mag}$.  This
modulation, whose period is typically a few minutes, has been detected 
only in the early and late phase of superoutbursts in some SU UMa-type
DNe (\cite{kat92}; \cite{nog98}; \cite{kat02}).    

\citet{ste82} discovered new variable stars in the course of their search 
for galaxies with a low-dispersion objective prism survey.  According to
\citet{ste82}, the ``No. 2'' object showed brightness variations of 
$V=15$--$17.5$.  In conjunction with strong Balmer emission lines
and a weak He~{\sc ii} 4686 \AA \ emission line, \citet{ste82} suggested
that it is a DN.  This object was called FBS 0756+164 by 
\citet{kop88} and given a general GCVS name of DW Cnc (\cite{kho81}).
\citet{kop88} supported this classification based on a new
spectrum whose characteristics are analogous to that reported by
\citet{ste82}.   

Here, we report on our time-series photometric monitoring of DW Cnc, in which
we discovered unprecedentedly large-amplitude and long-period QPOs.  In
the next section, we describe our observation method.  Our observation 
results are presented in section 3; we then propose possible
interpretations of them in section 4.  We summarize our findings in the 
final section.

\section{Observation}

\begin{table*}
\caption{Equipment of our observations.}
\label{tab:equ}
\begin{center}
\begin{tabular}{cccc}
\hline \hline
Site & Telescope & Camera & Filter \\
\hline
Kyoto & 25-cm Schmidt--Cassegrain & ST-7 & unfiltered \\
Ouda & 60-cm Ritchey--Chretien & PixelVision (SITe SI004AB chip) &
 $R_{\rm c}$\\
Nicholas Copernicus Observatory & 40-cm Newton & ST-7 &
 $I$ \\
Masaryk University & 60-cm Newton & ST-8 & $I$ \\
Erftstadt & 20-cm Schmidt--Cassegrain & ST-6B & unfiltered \\
\hline
\end{tabular}
\end{center}
\end{table*}

We performed CCD photometric observations at Kyoto, Ouda Observatory, 
Nicholas Copernicus Observatory and Planetarium, and Erftstadt.  
Our equipment and 
observation log are listed in tables \ref{tab:equ} and \ref{tab:log}, 
respectively.  After dark subtraction and flat fielding on our
obtained images, the differential magnitudes of the object were calculated 
with point spread function (PSF) photometry for images taken at 
Kyoto.  Aperture photometry was performed for images taken at the other 
sites.  The magnitude scales
of each observatory were adjusted to that of the Kyoto system, in which 
we used a comparison star, GSC 1363.2124.  The constancy of the
comparison brightness was checked by GSC 1363.2014.  We could obtain
magnitudes almost equal to the $R_{\rm c}$ system with an unfiltered ST-7 
camera, since its sensitivity peak is almost equal to that
of $R_{\rm c}$ system and the object has $B-V\sim 0$.  Heliocentric 
corrections to the observed times were applied before a following
analysis. 

\begin{table}
\caption{Journal of observations.}
\label{tab:log}
\begin{center}
\begin{tabular}{ccccc}
\hline \hline
Date (HJD) & Duration (hr) & $T_{\rm exp}$ (s) & $N$ & Site\\
\hline
2451986.9965 & 2.17 & 60 & 118 & O\\
2451987.9076 & 5.15 & 30 & 286 & O\\
2451988.2804 & 5.42 & 50 & 345 & N\\
2451988.4684 & 1.91 & 90 &  71 & E\\
2451988.9076 & 5.53 & 30 & 274 & O\\
2451988.9596 & 4.68 & 30 & 161 & K\\
2451990.0891 & 2.43 & 30 & 235 & K\\
2451991.9635 & 5.52 & 30 & 327 & K\\
2451995.0318 & 1.77 & 30 &  85 & K\\
2451996.3280 & 2.36 & 90 &  48 & E\\
2451999.3545 & 3.45 & 90 & 92 & E\\ 
2451999.9361 & 1.89 & 30 & 184 & K\\
2452000.9436 & 1.97 & 30 & 189 & K\\
2452001.9265 & 2.82 & 30 & 279 & K\\
2452003.2971 & 3.46 & 50 & 134 & M\\
2452004.0013 & 0.84 & 30 &  83 & K\\
2452004.2945 & 3.21 & 50 & 172 & N\\
2452005.0134 & 0.19 & 30 &  19 & K\\
2452006.0340 & 0.88 & 30 &  80 & K\\
2452009.9474 & 1.52 & 30 & 149 & K\\
2452011.9400 & 2.11 & 30 & 146 & K\\
2452012.3150 & 1.22 & 50 &  66 & N\\ 
2452014.9504 & 0.08 & 30 &   9 & K\\
2452015.9234 & 1.12 & 30 & 107 & K\\
2452016.9480 & 1.88 & 30 & 165 & K\\
2452018.9189 & 1.38 & 30 & 136 & K\\
2452019.9214 & 1.89 & 30 & 186 & K\\
2452021.9212 & 1.46 & 30 & 144 & K\\
2452022.9230 & 1.59 & 30 & 160 & K\\
2452029.9318 & 0.19 & 30 &  21 & K\\
2452032.9592 & 0.08 & 30 &  10 & K\\
2452033.9419 & 0.23 & 30 &  21 & K\\
2452039.0084 & 0.15 & 30 &  16 & K\\
2452039.9603 & 0.09 & 30 &  10 & O\\
2452040.9686 & 0.47 & 30 &  44 & K\\
2452041.9350 & 0.91 & 30 &  93 & O\\
2452041.9407 & 1.71 & 30 & 172 & K\\
2452045.9426 & 0.08 & 30 &  10 & K\\
2452046.9454 & 0.09 & 30 &  10 & K\\
2452047.9466 & 0.11 & 30 &  12 & K\\
\hline
\end{tabular}
\end{center}
{\footnotesize
Site: K = Kyoto, O = Ouda, N = Nicholas Copernicus Observatory,
 M=Masaryk Univesity Telescope, E = Erftstadt}
\end{table}

\section{Result}
\subsection{Long-Term Light Curve and Re-Classification of DW Cnc}

\begin{figure}
  \begin{center}
    \FigureFile(80mm,80mm){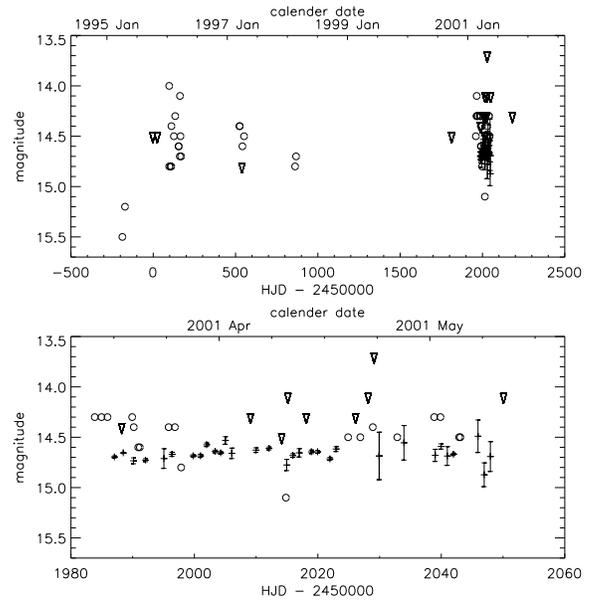}
  \end{center}
  \caption{Long-term light curve of DW Cnc.  The abscissa and the ordinate
 denote HJD (also calendar date) and the $R_{\rm c}$-magnitude, 
 respectively.  The open circles and the crosses are observations reported to 
 Variable Star Network (VSNET), and our CCD observations, respectively.  
 The open triangles show upper-limits of visual observations.  Upper panel: 
 Light curve including all visual observations from 1995 to 2001.  Lower 
 panel: Light curve focused on our CCD observations.}
\label{fig:lc1}
\end{figure}

Figure \ref{fig:lc1} shows long-term light curves of DW Cnc.  The crosses
with error bars represent the average magnitudes with standard errors 
each night.  The light curve in the upper panel includes all visual
observations reported to Variable Star Network 
(VSNET\footnote{$\langle$http://www.kusastro.kyoto-u.ac.jp/vsnet/$\rangle$}),  
which are 
depicted with the open circles and the triangles (upper-limits).  The lower
panel is a light curve focused on our CCD observations.  As can be
seen in the upper panel, the object varied between 14--15 mag 
most of the time.  During our observations, shown in the lower panel, the
object showed no significant fading or brightening trend.  We calculated 
the average magnitude to be $R_{\rm c}=14.68$.  

\citet{ste82} proposed that DW Cnc is a DN with the variation range of 
$V=15$--$17.5$.  The characteristic which we observed, as shown in
figure \ref{fig:lc1}, is, however, inconsistent with that of the DN light 
curves.  Z Cam stars, which form a sub-group of DNe, are known to
experience a phase called ``standstill'', which is a phase with constant
brightness interrupting ordinary outburst cycles (\cite{war95}).  
It is, however, less likely that the period of constant 
brightness of DW Cnc is a standstill, because the long monitoring of 
visual observations shows no ordinary outburst phase over six years.  
It may be possible that the long bright state is indeed a
quiescent state of a dwarf nova having a long recurrence time, such 
as WZ Sge stars.  This scenario, however, apparently contradicts the 
presence of the faint phase, which is indicated by a few observations fainter 
than 15 mag in figure \ref{fig:lc1} and the originally proposed
amplitude of $V=15$--$17.5$ (\cite{ste82}).

The long-term variations of DW Cnc are analogous to NLs or magnetic
CVs, rather than DNe.  Figure \ref{fig:lc1} indicates that the duration
of the possible faint state is probably shorter than that of the high
state.  Although we could obtain no estimation of the magnitude at the time
when the spectroscopy was performed by \citet{ste82} and \citet{kop88}, it is
less likely that both of the spectra were taken during the temporal 
short-duration faint states.  The case of a magnetic CV is thus
apparently inconsistent with the lack of a strong He~{\sc ii} emission 
line (\cite{ste82}; \cite{kop88}).  We hence suggest that DW Cnc is 
neither a DN nor a magnetic system, but a NL.  

We cannot conclude whether DW Cnc 
has changed its nature as a DN or it was just a mis-classification, since 
no light curve is available in \citet{ste82}.  On the other hand, the 
hints of possible faint states are reminiscent of the long-term 
variations of VY Scl-type stars, a sub-group of NLs showing temporary
low states (\cite{lea99}).  The variations between 14--15 mag seen in 
the visual observations can be explained by short-term variations with 
relatively large amplitudes, which we describe in the following section.  

\subsection{Discovery of Kilo-Second Quasi-Periodic Oscillations}

\begin{figure*}
  \begin{center}
    \FigureFile(180mm,110mm){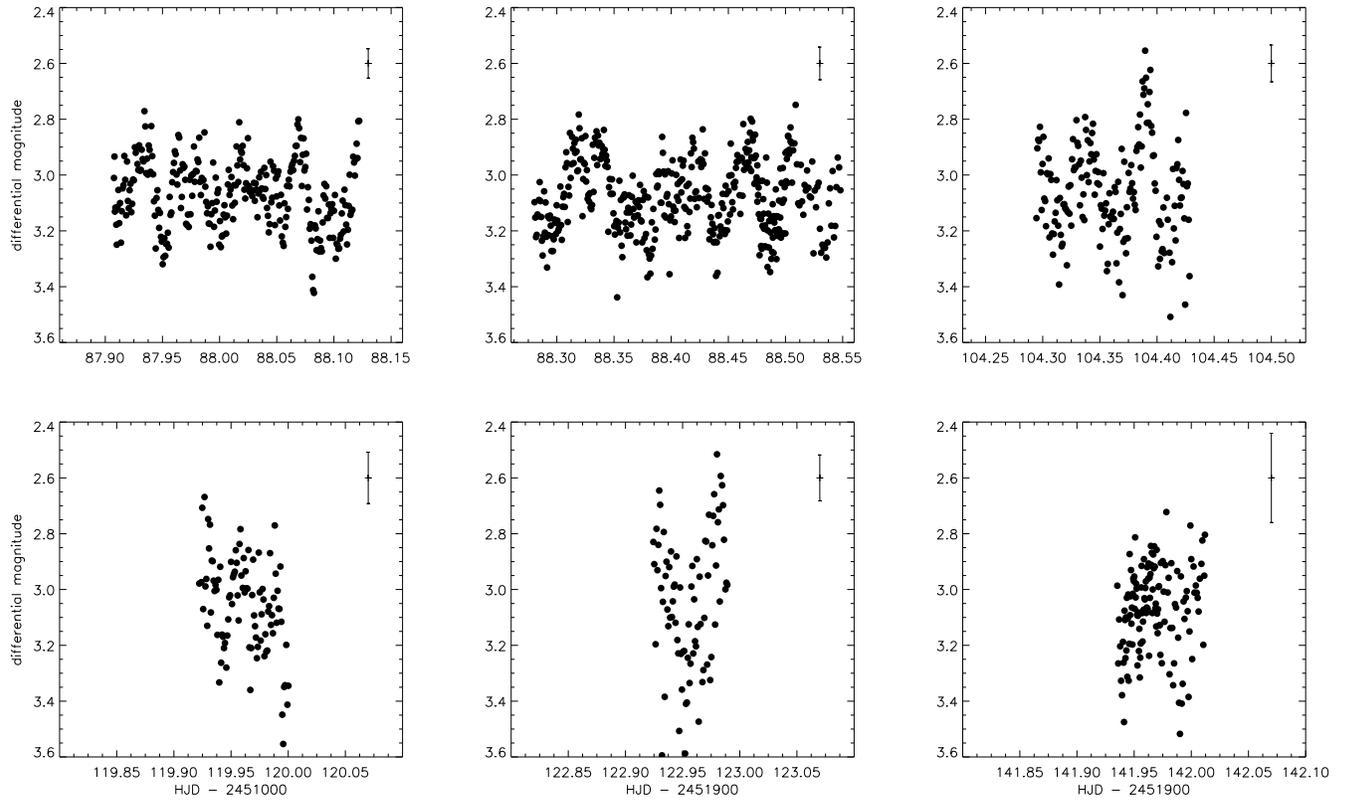}
  \end{center}
  \caption{Quasi-periodic variations of DW Cnc.  The abscissa and the ordinate
 denote HJD and the differential magnitude, respectively.  We show
 typical errors in each panel.}
\label{fig:lc2}
\end{figure*}

\begin{figure}
  \begin{center}
    \FigureFile(80mm,80mm){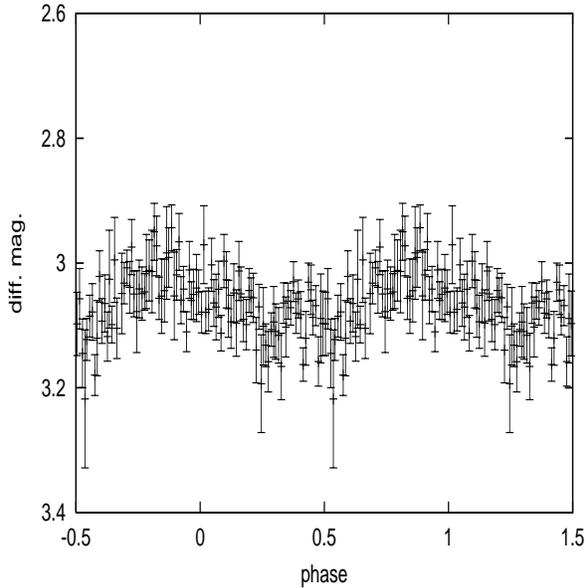}
  \end{center}
  \caption{Average light curve of modulations.  The abscissa and the ordinate
 denote the phase and the differential magnitude, respectively.  The
 phase was calculated with the 73.4-min period and an arbitrary epoch.}
\label{fig:ave}
\end{figure}

\begin{figure}
  \begin{center}
    \FigureFile(80mm,80mm){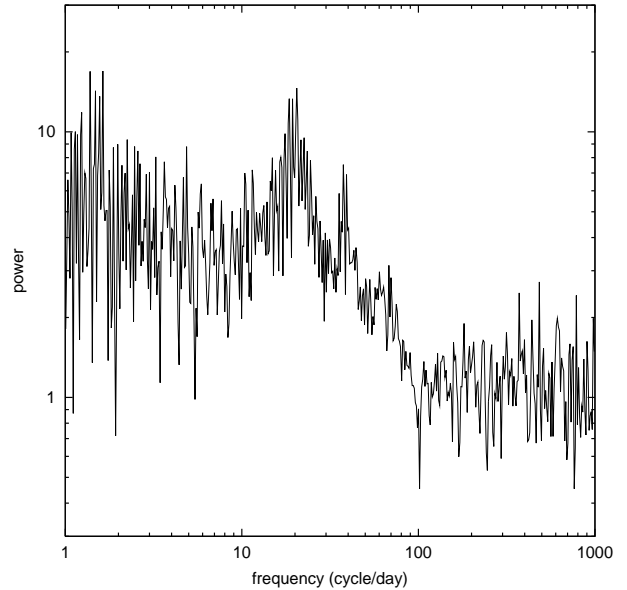}
  \end{center}
  \caption{Power spectrum of the light curve of DW Cnc.  The abscissa and
 the ordinate denote the frequency in cycles per day and the power in
 arbitrary units, respectively.}
\label{fig:pow}
\end{figure}

During our CCD monitoring, we detected oscillations whose typical light 
curves are shown in figure \ref{fig:lc2}.  As shown in this figure, they 
appear to have no constant period, but variable profiles and 
a quasi-periodic nature.  They have a coherence time of at least a few
cycles, as can be seen in figure \ref{fig:lc2}.  We cannot find any firm
evidence of a long-term coherence of these variations within our
limited available data due to the short base-lines of each observation.  
Through a period analysis, we confirmed that no clear 
coherent peak exists in the 
power spectrum between periods of 1--200 min.  
We show a wide-range power spectrum in figure \ref{fig:pow}.  In this
figure, the abscissa and the ordinate denote the frequency in cycle
day$^{-1}$ 
and the power in arbitrary units, respectively.  We can find two prominent
peaks with wide widths, which demonstrate the QPO nature of the 
oscillations.  We determined their center periods to be $37.5\pm 0.1$ and 
$73.4\pm 0.4$ min by fitting Gaussian functions.  The shorter QPO period 
is probably the first harmonic of the longer one; however, due to the
small difference between them, we cannot completely exclude the
possibility that they are periodicity independent.  We show the average
light curve of modulations folded with the 73.4-min period in figure 
\ref{fig:ave}.  We can see the main peak around the phase $\sim 0.8$ and
the sub peak around the phase $\sim 0.4$.  

The QPO in DW Cnc is unique regarding its long lifetime, large 
amplitude and long period of kilo-second.  QPOs in CVs generally
have periods of over one order of magnitude shorter than that in DW Cnc 
(\cite{pat81a}).  Although kilo-second QPOs have been reported in some CVs
(e.g. \cite{uda88}; \cite{kra99}), their amplitudes are quite small 
($\sim$ a few percent) compared with those in DW Cnc.  
The amplitude of QPOs in
DW Cnc is typically about 0.3 mag and, furthermore, we detected a large 
flare-like peak whose amplitude reached to 0.6 mag, as can be seen in the
upper-right panel of figure \ref{fig:lc2}.  We clearly detected the 
oscillation, or a part of it, in our all observations with a long
time-coverage, as shown in figure \ref{fig:lc2}.  Although the light
curves in the lower-middle and right panel in figure \ref{fig:lc2} are 
more noisy due to the bad seasonal condition, they show a clear presence of
oscillations with amplitudes and period similar to those in the upper
panels.  It is thus probable that the QPO has appeared throughout our 
observations for 61 days.  

Except for the two QPOs, the power spectrum shows other noteworthy 
characteristics.  It indicates the presence of a significant power for   
frequencies lower than 10 ${\rm cycle}\; {\rm d}^{-1}$.  The power
suddenly weakens at frequencies higher than the 37.5-min QPO.  These 
features of the power spectrum are qualitatively analogous to that in
black hole binaries (\cite{rut99}).  The poisson noise dominates for  
frequencies higher than 100 cycle d$^{-1}$, where the spectrum becomes
flat. 

\section{Discussion}

In this section, we compare DW Cnc with the other CVs
which also show similar oscillations to evaluate whether their proposed 
models can be applied to the QPO in DW Cnc.

\subsection{Spin of White Dwarf}

Some IPs show periodic variations on the order of kilo-second
(e.g. \cite{sch85}; \cite{pat81b}), which are interpreted as being the 
rotational 
period of a white dwarf.  In the case of GK Per, the long-period QPO 
($\sim 5000\;{\rm s}$) reported in \citet{wat85} is proposed to be a
beat phenomenon between the spin of a white dwarf and blobs orbiting at
a disk-overflow impact site (\cite{hel94}) or between 
the accreting curtain and 
ordinary DN QPOs (\cite{mor99}).  The long period seen in DW Cnc
may prefer a beat scenario rather than a direct spin variation.
The most essential objection against this model is that DW Cnc shows no 
direct evidence for an IP.  The two spectra obtained by \citet{ste82} and 
\citet{kop88} show no strong He~{\sc ii} emission and, moreover, we detected no
firm evidence of a short periodicity originating from the white-dwarf spin.  
It may be possible that the shorter periodicity of 37.5-min is
not the first harmonic of the longer periodicity, but is associated with
the spin period.  In this case, the kilo-second QPO can be interpreted
with the beat between the white-dwarf spin and $\sim 25$-min QPO.  
Within our available data, the broad profile of the 37.5-min peak in 
the power spectrum, however, supports that this also has the QPO nature,
and probably corresponds to the first harmonic of the 73.4-min QPO.  

\subsection{Disk Oscillation}

Thin-disk oscillations have been proposed to explain the characteristics of
QPOs both in DNe and X-ray binaries (\cite{kat01}).
In a standard disk during DN outbursts, the oscillation wave 
propagating inward is reflected at the radius where the wave frequency 
is equal to the epicyclic frequency.  Since the wave cannot propagate 
in an outer cool disk, the wave is trapped in a narrow region.  
The wave is thus excited enough to be detected as QPOs in DNe (\cite{yam95};
\cite{yam96}).  Although DW Cnc is definitely not a DN, we can expect
the condition of an accretion disk analogous to that during DN outbursts, 
since the object is most likely to be a NL, which is generally believed
to have a sufficiently high mass-accretion rate to form a
fully-ionized disk (\cite{osa96}).  The  
Balmer emission lines, furthermore, indicate the presence of an optically
thin plasma around the disk.  In the case of DN outbursts, the QPO
frequency is theoretically expected to become higher with time due to
an inward propagation of the transition layer (\cite{yam95}).  
On the other hand, the model can explain the long
lifetime of the QPO in DW Cnc with no apparent period shortening, since
we can expect a quasi-steady disk system in the case of NLs.
\citet{kra99} reported 50-min QPOs in a NL, MV Lyr and also suggest
applying of the trapped disk oscillation model. 

According to \citet{yam95}, the trapped oscillation model provides the  
expected QPO period, which is the inverse of the Keplerian frequency at
the transition layer, that is, $P_{\rm QPO}\sim 600\;{\rm s}\; \times 
(r_{\rm front}/10^{10}\;{\rm cm})^{3/2}(M/M_\odot)^{-1/2}$.  With the
$\sim 73\;{\rm min}$ QPO, we estimate $r_{\rm front}=3.8\times
10^{10}\;{\rm cm}$, assuming a white-dwarf mass of $1M_{\rm solar}$. 
This $r_{\rm front}$ is reasonable in CVs with an orbital period longer 
than about three hours.  The disk-oscillation scenario can thus be an
acceptable model for the QPO in DW Cnc if its orbital period is above 
the period gap.  This scenario, however, was originally proposed to
explain the ordinary DN QPO, which has amplitudes typically smaller
than $0.1\;{\rm mag}$ (e.g. \cite{yam95}).  While \citet{kra99}
proposed this model for $\sim 0.2\;{\rm mag}$ QPOs observed in MV Lyr, 
the amplitude reaching $\sim 0.3\;{\rm mag}$ in DW Cnc may be too large 
for this model to reproduce.   

\subsection{Superhump}

The large amplitude and the relatively long period are rather analogous 
to superhumps which are observed during superoutbursts in SU UMa-type
DNe (\cite{war85}).  Superhumps are modulations which generally have a 
period a few percent longer than an orbital period.  According to the
tidal instability model, this phenomenon is caused
by the gradual precession of an eccentric disk which is developed by 
an enhanced tidal effect at the 3:1 resonance (\cite{osa89}).  
As expected from this model, superhumps
have been reported not only in SU UMa-type DNe, but also in NLs and AM
CVn stars (\cite{pat97}; \cite{tay98}).  If QPOs in DW Cnc are
superhumps, its period is probably the longer one, i.e., 73.4 min.  
This is because the 37.5-min period indicates that the system is an 
AM CVn object; however,
it contradicts the Balmer emissions observed in the optical spectra.  
According to \citet{pat01}, a system with a shorter orbital period 
generally has a smaller superhump excess, $\varepsilon = |P_{\rm
superhump}-P_{\rm orb}|/P_{\rm orb}$.  Assuming $0.02< \varepsilon
<0.03$ (\cite{pat01}), we can roughly estimate the orbital period of 
DW Cnc to be 71.3--72.0 min.  This is the shortest orbital period in 
hydrogen-rich CVs, except for the SU UMa-type DNe, V485 Cen 
($P\sim 59 \; {\rm min}$; \cite{aug96}) and 1RXS J232953.9+062814 
($P\sim 64 \; {\rm min}$; \cite{uem01}).  DW Cnc may thus be a permanent 
superhumper below the period minimum if its orbital period is 70--80
min.  

\section{Summary}

Our photometric monitoring of the cataclysmic variable DW Cnc revealed
that the object remained at $R_{\rm c}=14.68\; {\rm mag}$ for 61 days.  
This characteristic of the long term light curve indicates that it is
not a dwarf nova, which was originally proposed in \citet{ste82}, but 
a nova-like object.  It is less likely that the object
includes a white dwarf with a strong magnetic field, since both optical
spectra obtained by \citet{ste82} and \citet{kop88} show only weak
He~{\sc ii} emission.  
The possible presence of faint states implies the VY Scl-type nature of
DW Cnc.  We discovered kilo-second quasi-periodic oscillations (QPOs) in 
DW Cnc, whose center periods were calculated to be $37.5\pm 0.1$ and
$73.4\pm 0.4\; {\rm min}$.  Compared with typical dwarf-nova QPOs, the
oscillation of DW Cnc is unique regarding the large amplitude, 
long period, and long lifetime.  We propose possible interpretations for 
DW Cnc QPOs: (i) The trapped disk oscillation scenario, which can be
acceptable if the orbital period is above the period gap. (ii) The 
superhump scenario, which can be acceptable if it is below the period 
minimum.  

\vskip 3mm

We are grateful to many amateur observers for supplying their vital
visual CCD estimates via VSNET.  Part of this work was supported by a 
Research Fellowship of the Japan Society for the Promotion of Science 
for Young Scientists (MU).  We would like to thank to 
Masaryk University for giving an observing time.


\end{document}